\documentclass{PoS}
\usepackage{wrapfig,rotating}

\title{
The effects of combined HERA and recent Tevatron $W\to\ell\nu$ charge asymmetry
data on the MSTW PDFs}

\ShortTitle{Effect of new data on MSTW PDFs}

\author{\speaker{R.S. Thorne},$^a$ A.D. Martin$^b$, W.J. Stirling$^c$
and G. Watt$^d$\\
 \llap{$^a$}Department of Physics and Astronomy, University College London, WC1E 6BT, UK\\ 
 \llap{$^b$}Institute for Particle Physics Phenomenology, University of Durham, DH1 3LE, UK\\
 \llap{$^c$}Cavendish Laboratory, University of Cambridge, CB3 0HE, UK\\
 \llap{$^d$}Theory Group, Physics Department, CERN, CH-1211 Geneva 23, Switzerland\\ 
        E-mail: \email{thorne@hep.ucl.ac.uk}, \email{A.D.Martin@durham.ac.uk},
      \email{wjs2@cam.ac.uk}, \email{Graeme.Watt@cern.ch}}

\abstract{We examine the effect of including the `combined' HERA structure 
function data in the MSTW global fit for parton distribution functions (PDFs).  
The combined neutral-current HERA data have a significant,
if not dramatic, effect, of up to 2--3\% at NLO for $Z$ boson and Higgs production
at the Tevatron and LHC,
and a generally slightly smaller effect, particularly on LHC processes, at NNLO.  This is an
amount comparable, or less than, the typical PDF uncertainties, and hence we do not intend
to release an imminent update to the MSTW 2008 fit.
We also investigate the consistency with the recent D{\O} data on electron and muon charge
asymmetry from $W$ decays and the direct CDF measurement of the $W$ charge asymmetry.
The D{\O} lepton charge asymmetry data imply a fairly large change to the
down-quark distribution and/or large 
nuclear corrections to be applied when fitting to deuterium structure function data, while the CDF 
$W$ charge asymmetry data are more consistent with the existing PDFs. 
However, it is difficult to reconcile all of the Tevatron $W\to\ell\nu$ charge asymmetry data sets
with the fit, and to some extent, with each other.}

\FullConference{XVIII International Workshop on Deep-Inelastic Scattering and Related Subjects, DIS 2010\\
		April 19-23, 2010\\
		Firenze, Italy}

\begin{document}

\section{HERA combined data}

The MSTW 2008 global fit~\cite{Martin:2009iq} for PDFs 
contains a very large variety of data sets, including a 
number from H1 and ZEUS on structure functions. These structure function 
data (along with some newer sets) have recently been 
combined in~\cite{Aaron:2009wt}. The increase in precision comes about not only 
from a combination of statistics, but from the fact that one collaboration often 
controls a source of systematic uncertainty better than the other, so the
systematic uncertainties can be greatly reduced by combination. In addition,
this treatment of the correlated errors means that the central values are not 
simply the weighted averages. These data were fit in~\cite{Aaron:2009wt} and 
differences between previous H1 and ZEUS fits noted. 
Here we, too, investigate the inclusion of the combined neutral-current (NC) data 
instead of the component sets, adding the errors in quadrature for the moment 
(systematics in the combined data set are small). We include the uncombined
charged-current data which are statistics dominated. We fit to data with 
$Q^2\geq 2$~GeV$^2$ (553 pts.), but also look at results for 
$Q^2\geq 3.5$~GeV$^2$ (524 pts.) to 
compare with~\cite{Aaron:2009wt}. All other details of the fit are as in~\cite{Martin:2009iq}.

\begin{wrapfigure}{r}{0.5\columnwidth}
\vspace{0.4cm}
\centerline{\includegraphics[width=0.43\textwidth]{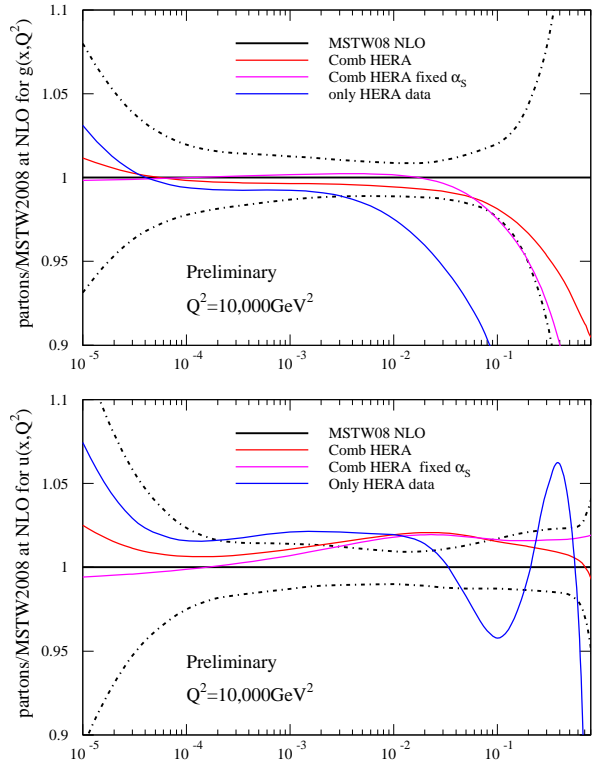}}
\caption{The ratio of NLO PDFs fit to the 
combined HERA data to the default MSTW08 distributions.}
\vspace{-0.5cm}
\label{partratios} 
\end{wrapfigure}

At NLO the fit quality achieved is $2610/2471$ for the total data. For the HERA
NC data it is $600/553$ and $530/524$, compared to $483/524$ in~\cite{Aaron:2009wt}.
In order to check if the worse quality is due to the other data in the MSTW fit we also
fit \emph{only} to HERA structure function data. This results in $515/553$ and 
$445/524$, now much better than
the HERA fit~\cite{Aaron:2009wt}, presumably due to extra parameterisation freedom. Clearly 
the fit quality is significantly affected by tension with other data sets. The MSTW global fit with HERA 
combined data requires $\alpha_S(M_Z^2)=0.1215$, a little higher than the 
default 0.1202. Keeping at the default value results in a fit quality 10 
worse for HERA data and the global fit. Our `HERA data only' fit gives 
$\alpha_S(M_Z^2)=0.123$. The resulting PDFs are shown in Fig.~\ref{partratios}
as a ratio to MSTW08 with the $1\sigma$ uncertainty. For
the global fits the up quark strays outside this uncertainty at
$x\sim 0.01$, otherwise there is little change. 
The predictions for hadron collider processes give 2--3\% variations for $Z$ production, 
but less for Higgs production (from gluon--gluon fusion, with $M_H=120$~GeV), see Table~\ref{cstable}.    
Our fit to `HERA data only'
gives a similar small-$x$ gluon but for $x>0.01$ it is far softer, while the up-quark distribution
is not well constrained in shape for $x>0.01$. These PDFs compare 
very badly to many unfitted data sets. Both the variation and comparison to 
unfitted data are far worse than for the PDFs in \cite{Aaron:2009wt}, showing that it is 
implicit constraints on the quark parameterisation rather than a real 
data constraint that render these PDFs similar to those from global fits at 
high $x$.

\begin{wraptable}{r}{0.69\columnwidth}
\vspace{-0.0cm}
\begin{tabular}{|l|ll|ll|}
\hline
PDF set & $B_{l^+l^-}\!\!\cdot\!\sigma_Z$(nb)   & $\!\!\!\sigma_H$(pb)$\!$ & $B_{l^+l^-}\!\!\cdot \!\sigma_Z$(nb)& $\!\!\!\sigma_H$(pb)$\!$\\
\hline
NLO& Tevatron&&LHC $(14~{\rm TeV})\!\!\!\!\!\!\!\!$&\\
\hline
     MSTW08 & $0.243^{+2.4\%}_{-2.0\%}$ &$\!\!\!0.746^{+5.0\%}_{-4.4\%} \!\!\!$& $2.00^{+2.1\%}_{-1.8\%}$ &$\!\!\! 40.7^{+3.0\%}_{-2.7\%} \!\!\!$ \\  
     New ${\rm HERA} \!\!\!$& $+3.1$\% & $\!\!\!-0.7\%\!$ & $+ 2.5\% $& $\!\!\!+1.2\% \!$ \\    
fix $\alpha_S(M_Z^2)$ & $+3.0\%$ & $\!\!\!-4.0\%\!$ & $+1.8\%$ & $\!\!\!-0.8\%\!$\\        
\hline
NNLO&Tevatron&&LHC $(14~{\rm TeV})\!\!\!\!\!\!\!\!$&\\
\hline
     MSTW08 & $0.251^{+2.2\%}_{-1.8\%}$ & $\!\!\!0.955^{+5.4\%}_{-4.8\%} \!\!\!$& $2.05^{+2.6\%}_{-2.1\%}$ & $\!\!\!50.5^{+3.6\%}_{-2.7\%}\!\!\!$  \\   
     New ${\rm HERA} \!\!\!$& $+3.0\%$ & $\!\!\!+0.2\%\!$& $+1.8\%$ & $\!\!\!+0.8\%\!$ \\    
fix $\alpha_S(M_Z^2)$ & $+2.6\%$ & $\!\!\!-2.9\%\!$ & $+1.2\%$ &$\!\!\!-1.0\%\!$ \\    
\hline
    \end{tabular}
\vspace{-0.2cm}
\caption{The cross sections at the Tevatron and LHC with ``PDF+$\alpha_S$'' uncertainties 
and the changes in fits using the new combined HERA data.} 
\vspace{-0.4cm}
\label{cstable}
\end{wraptable}

At NNLO the fit quality for the HERA
NC data is $575/553$ and $529/524$, better than NLO. 
$\alpha_S(M_Z^2)$ moves only from 
0.1171 to 0.1181, and fixing it leads to $\chi^2 \approx 8$ higher. Fitting only 
HERA data gives $494/553$ and $467/524$. The change in PDFs is similar to NLO, but with a tendency to dip slightly at $x<0.001$. The NNLO predictions  
have generally less variation at the LHC than at NLO, see Table~\ref{cstable}. 
The effect of the combined data is significant, and will be included in an
updated set soon, but certainly is not dramatic enough to invalidate the 
present MSTW 2008 PDFs~\cite{Martin:2009iq}. 

\vspace{-0.3cm}

\section{Tevatron lepton charge asymmetry from $W$ decays}

\vspace{-0.25cm}

There are new D{\O} data on electron~\cite{Abazov:2008qv} and 
muon~\cite{D0muon} charge asymmetry and CDF $W$ charge asymmetry
data~\cite{Aaltonen:2009ta}. These are far more precise than the previous 
measurements~\cite{Acosta:2005ud,Abazov:2007pm}, which 
already give the main constraint on some PDF eigenvectors. 
These new data should constrain 
the down quark for $0.01<x<0.7$, where the current 
main constraint is deuterium DIS
and is subject to uncertainty from nuclear corrections (a source not
included in PDF uncertainties). The current fit to 
asymmetry data is moderate~\cite{Martin:2009iq}.  
At NLO  $\chi^2 = 25/10$ (D{\O})~\cite{Abazov:2007pm} and $\chi^2 = 29/22$
(CDF)~\cite{Acosta:2005ud}. 

\begin{wraptable}{r}{0.69\columnwidth}
\vspace{-0.3cm}
\hspace{-0.1cm}
\begin{tabular}{|l|l|l|l|l|l|}
\hline
    fit       &  $\chi^2 \!/\!12\!$ {$(e)\!\!\!\!$} &   
$\chi^2 \!/\!12\!$ {$(e)\!\!\!\!$} &   $\chi^2 \!/\!12\!$ {$(e)\!\!\!\!$} & 
$ \chi^2\!/\!2689\!\!\!$ &   $ \chi^2 \!/\!16\!$ {$(\mu)\!\!\!\!$}         \\
 $p_T$  (GeV)    &  $ > 25$  &   25--35 
&   35--45 & non-{ D{\O}}$\!$ &  two bins
        \\
\hline
{Standard} & & & & &\\
{2008 data}  & {116}  & {19} & {144} & {2518} & 
{542}   \\
{D{\O}$_{e}$}   & 71  &  {\bf 23}  &  {\bf 81} & 2551 & 358\\
{D{\O}$_{e}\!$} (w)  & 25  &  {\bf 10}  &  {\bf 23} & 2942 & 183 \\
{D{\O}$_{\mu}$}  & 26  &  55  &  58 & 2640 & {\bf 119}\\
{D{\O}$_{\mu}\!$} (w)  &  33   &  79  & 88  & 3131 & {\bf 10}  \\
{D{\O}$_{\mu}\!$}  ${\rm cut} \!\!\!$ & 33 & 52  & 55  & 3190 & {\bf 26} \\
\hline
{Deut. Corr.} & & & & & \\
{2008 data}  & {25}  &  {32}  &  {42} & {2455}
 & {140} \\
{D{\O}$_{e}\!$} (w)  & 25  &  {\bf 9}  & {\bf 23}  & 2551 & 192\\
{D{\O}$_{\mu}\!$} (w)  & 38  &  67  & 75   &  2649 & {\bf 11} \\
{D{\O}$_{e+\mu}\!$} (w)  & 24  &  {\bf 16}  & {\bf 40}   &  2848 & {\bf 42} \\
{D{\O}$_{e}$} $p_T\!\!>\!\!25\!\!\!$  & {\bf 23}  &  38  & 32   & 2454 & 229 \\
\hline
   \end{tabular}
\vspace{-0.2cm}
\caption{Description of D{\O} lepton asymmetry and  
remaining data, without (upper) and with (lower) deuterium corrections. 
(w) denotes a high weight. Asymmetry data included in each fit are shown in bold type.} 
\vspace{-0.5cm}
\label{asymmtable}
\end{wraptable}

\begin{wrapfigure}{r}{0.6\columnwidth}
\centerline{\includegraphics[width=0.66\textwidth]{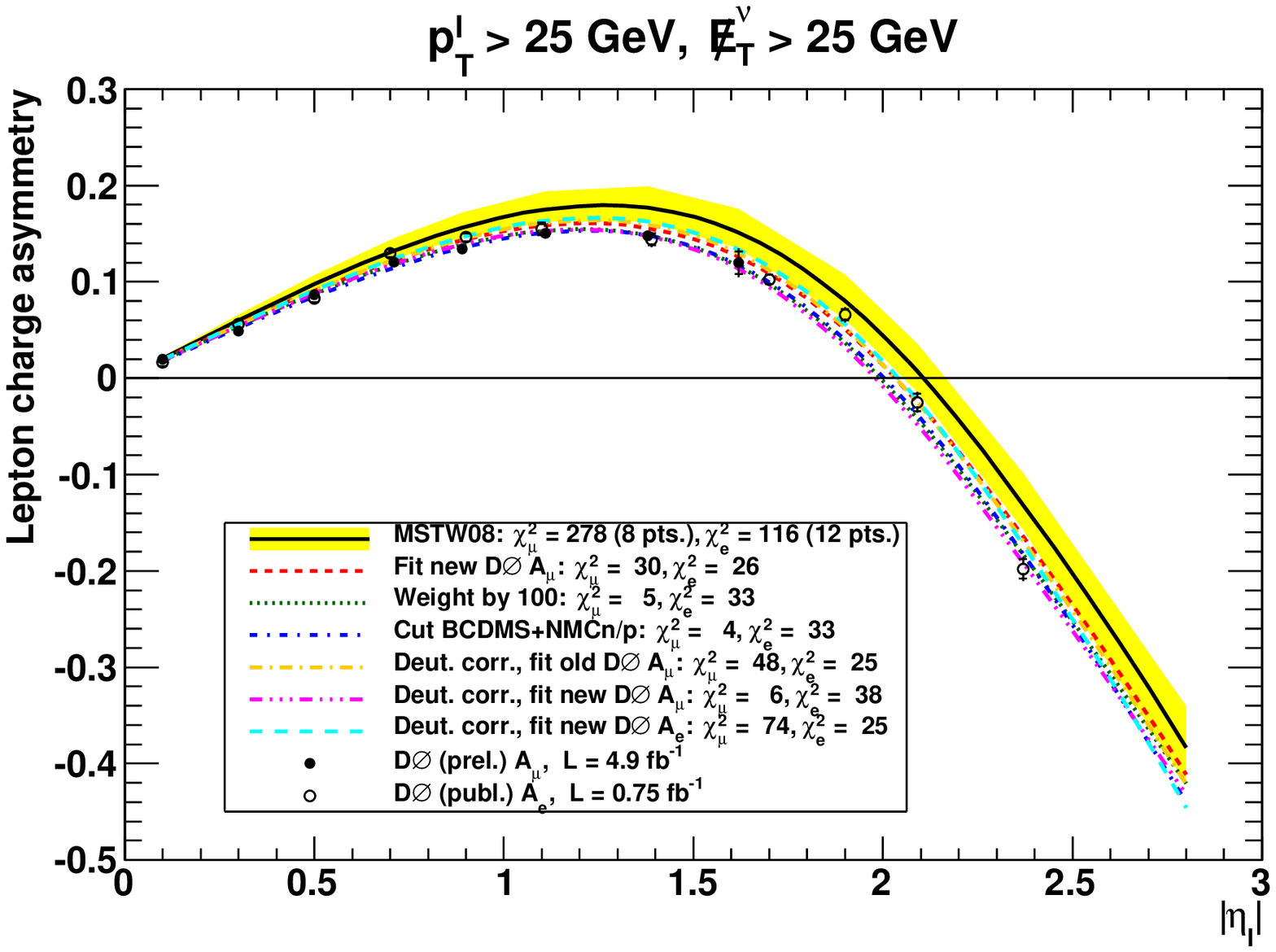}}
\centerline{\includegraphics[width=0.66\textwidth]{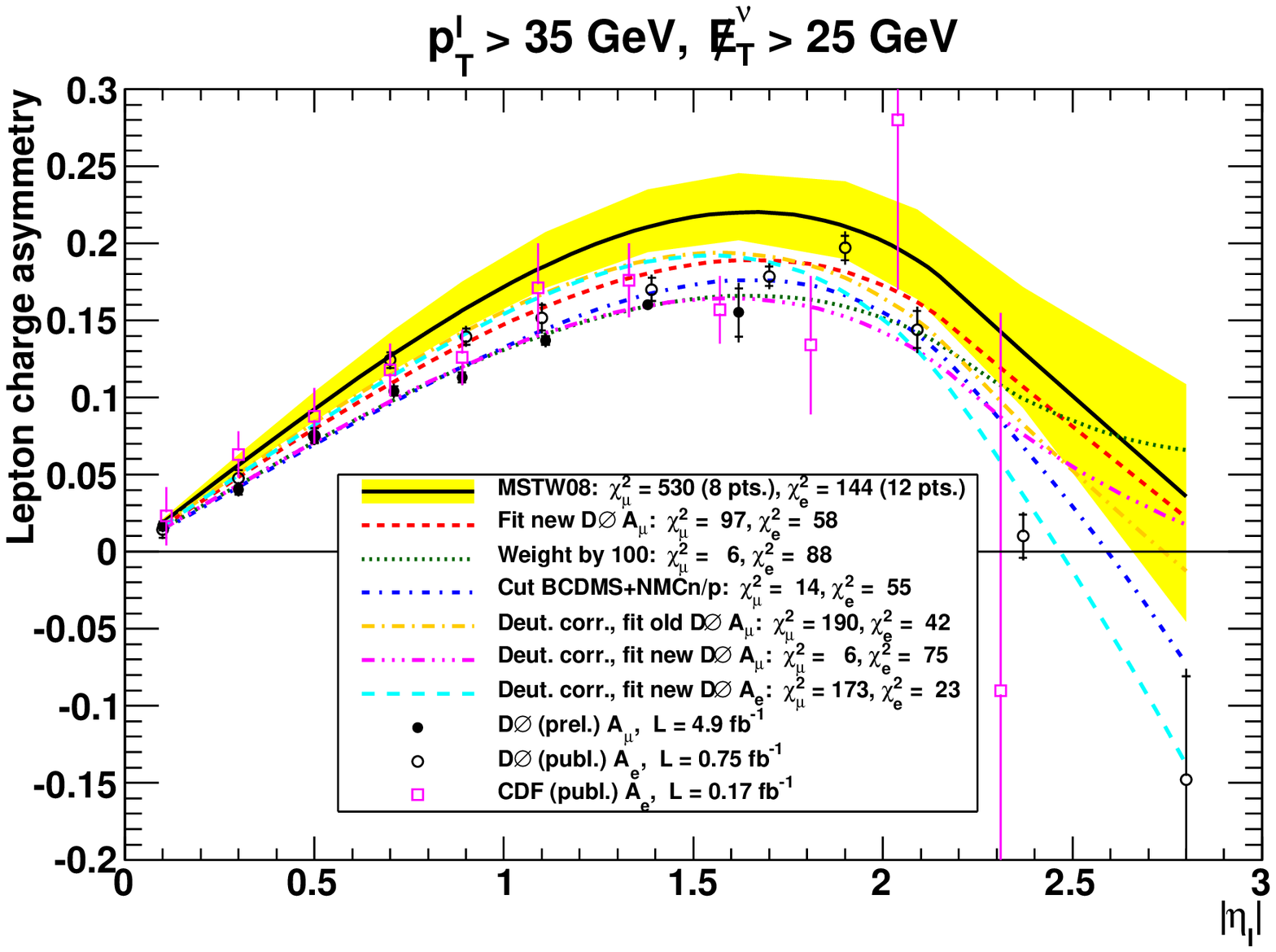}}
\vspace{-0.4cm}
\caption{The fit to D{\O} lepton asymmetry data for all $p_T>25$~GeV (top)
and 35 GeV $<p_T<$ 45~GeV (bottom).}
\vspace{-0.4cm}
\label{asymmfit} 
\end{wrapfigure}

The new data cause worse problems.  
Standard MSTW fits give a very poor comparison to  both D{\O} electron
and muon data, as seen in Table~\ref{asymmtable} and 
Fig.~\ref{asymmfit}. 
We have tried a variety of alternatives, 
weighting the asymmetry data and/or making cuts on other data in 
the fit. The fit quality to D{\O} $e$ and $\mu$ data and other data 
for these variations  is also 
shown. Here `cut' means the omission of BCDMS proton 
and deuterium data
and NMC $n/p$ data which are very 
badly described. 
NNLO corrections \cite{Melnikov:2006kv,Catani:2009sm,
Catani:2010en} do help the fit, but only marginally.

\begin{wrapfigure}{r}{0.40\columnwidth}
\vspace{-0.9cm}
\centerline{\hspace{-0.3cm}\includegraphics[width=0.35\textwidth]{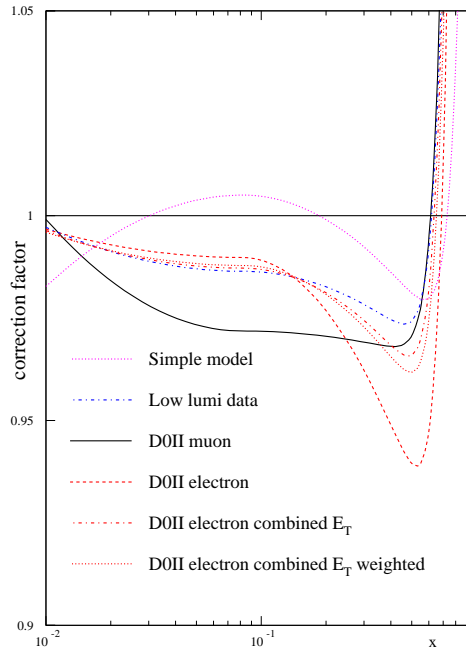}}
\caption{The deuterium corrections.}
\vspace{-0.5cm}
\label{deutcorrs} 
\end{wrapfigure}

In order to try to improve the fit, 
an extra parameter was added to both the high-$x$ valence 
quark distributions. This had negligible impact in the fit quality and 
extracted PDFs. Then variations in the deuterium corrections
to the structure function data fit were tried. In the standard MSTW fit these 
data have small corrections for shadowing at small-$x$, but
none at high-$x$. Removing these corrections and refitting 
using the standard MSTW08 data leads to $\chi^2 = 19/10$ (D{\O}) and 
$\chi^2 = 25/22$ (CDF), a significant improvement. 
The up and down quarks change by 1--2\% near $x=0.02$. 
Given this mild success we also tried a more sophisticated approach to 
corrections for deuterium data, i.e.~$Q^2$-independent
deuterium corrections for all $x$ applied to theory 
by means of a smooth function with $4$ free parameters.
This improves the quality of the fit to non-asymmetry data significantly,
as seen in the lower half of Table~\ref{asymmtable}, 
and using the standard MSTW08 data sets gives $\chi^2 = 6/10$ (D{\O})
and $\chi^2 = 21/22$ (CDF). When fitting to the newer 
D{\O} asymmetry data the free deuterium corrections also help a great deal,
as seen in Table~\ref{asymmtable} and Fig.~\ref{asymmfit}. 
A good fit (given the scatter of data) can be found for the electron data in the combined $p_T$ bin, with no
deterioration in fit quality for other data types. However, although very 
good fits to the electron data in separate-$p_T$ bins, or to the muon data, can be 
found, they both, especially the latter, result in a deterioration in the fit 
quality to other data. We also find that the muon data and electron 
data cannot be fit near their best individual quality simultaneously, see 
Table~\ref{asymmtable}. The required deuterium 
corrections are shown in Fig.~\ref{deutcorrs}. The general shape is similar 
to expectations, labelled ``simple model'', but in all cases,  especially when 
fitting muon data, the correction is low in the region of $x=0.1$. 
The main change in the PDFs, when deuterium corrections are applied, is 
an increase in the $d(x,Q^2)$ for $x>0.02$, and, in particular, 
a $10\%$ increase for $x\approx 0.4$ at $Q^2 = 10^4$~GeV$^2$.

As well as tensions  between the two D{\O} data sets there are also conflicts 
with the quality of the comparison to CDF $W$ asymmetry data 
\cite{Aaltonen:2009ta}.
The MSTW08 PDFs give (an approximate) $\chi^2\approx 28/13$ which, 
given the scatter of points, is 
quite good. The MSTW fit to the standard data with deuterium corrections 
gives $\chi^2\approx 24/13$, and as seen this also fits the combined-$p_T$
D{\O} electron data well. Good fits to the separate $p_T$ electron data and/or 
muon data give $\chi^2 > 100$ for comparison to the CDF $W$ asymmetry data,
due to too much asymmetry.

We conclude from 
this study that inclusion of deuterium corrections and an investigation
of their uncertainty is important for global fits. The deuterium data can 
be fit without them, but they improve the comparison even with the older low
statistics asymmetry data. An examination of the more recent asymmetry data
leads us to conclude that the maximally consistent sets in the fit are the CDF 
$W$ asymmetry data and the combined-$p_T$ D{\O} electron data. 
However, a good fit to the latter, without seriously affecting the 
rest of the global fit, requires slightly unlikely deuterium corrections.
In summary, it seems at 
present that the difficulties in reconciling the different asymmetry data sets
with theory, and with each other, necessitates further study, both of data
and the theory to be applied, before the true impact
in a global fit can be understood.


\vspace{0.1cm}

\begin{thebibliography}{99}


\vspace{-0.0cm}


\bibitem{Martin:2009iq}
  A.~D.~Martin, W.~J.~Stirling, R.~S.~Thorne and G.~Watt,
  \emph{Parton distributions for the LHC},
  {\it Eur.\ Phys.\ J.\ } C {\bf 63} (2009) 189
  [{\tt arXiv:0901.0002}].

\bibitem{Aaron:2009wt}
  F.~D.~Aaron {\it et al.}  [H1 and ZEUS Collaborations],
  \emph{Combined Measurement and QCD Analysis of the Inclusive $ep$ Scattering Cross
  Sections at HERA,}
  {\it  JHEP} {\bf 1001} (2010) 109
  [{\tt arXiv:0911.0884}].

\bibitem{Abazov:2008qv}
  V.~M.~Abazov {\it et al.}  [D{\O} Collaboration],
  \emph{Measurement of the electron charge asymmetry in $p \bar p \to W+X\to e \nu+X$ events at
  $\sqrt{s}=1.96$ TeV},
  {\it Phys.\ Rev.\ Lett.\ } {\bf 101} (2008) 211801
  [{\tt arXiv:0807.3367}].

\bibitem{D0muon}
D{\O} Collaboration, \emph{Measurement of the muon charge asymmetry in $p \bar p \to W+X\to\mu\nu+X$ 
events using the D{\O} detector}, D{\O} Note 5976-CONF.  

\bibitem{Aaltonen:2009ta}
  T.~Aaltonen {\it et al.}  [CDF Collaboration],
  \emph{Direct Measurement of the $W$ Production Charge Asymmetry in $p\bar{p}$
  Collisions at $\sqrt{s} = 1.96$ TeV,}
 {\it  Phys.\ Rev.\ Lett.\ } {\bf 102} (2009) 181801
  [{\tt arXiv:0901.2169}].

\bibitem{Acosta:2005ud}
  D.~E.~Acosta {\it et al.}  [CDF Collaboration],
  \emph{Measurement of the forward-backward charge asymmetry from $W \to e \nu$
  production in $p\bar{p}$ collisions at $\sqrt{s} = 1.96$ TeV,}
  {\it Phys.\ Rev.}  D {\bf 71} (2005) 051104
  [{\tt hep-ex/0501023}].

\bibitem{Abazov:2007pm}
  V.~M.~Abazov {\it et al.}  [D{\O} Collaboration],
  \emph{Measurement of the muon charge asymmetry from $W$ boson decays,}
  {\it  Phys.\ Rev.\ } D {\bf 77} (2008) 011106
  [{\tt arXiv:0709.4254}].

\bibitem{Melnikov:2006kv}
  K.~Melnikov and F.~Petriello,
  \emph{Electroweak gauge boson production at hadron colliders through
$\mathcal{O}(\alpha_S^2)$,}
  {\it Phys.\ Rev.\ } D {\bf 74} (2006) 114017
  [{\tt hep-ph/0609070}].

\bibitem{Catani:2009sm}
  S.~Catani {\it et al.},
  \emph{Vector boson production at hadron colliders: a fully exclusive QCD
calculation at NNLO,}
  {\it Phys.\ Rev.\ Lett.\ } {\bf 103} (2009) 082001 
  [{\tt arXiv:0903.2120}].

\bibitem{Catani:2010en}
  S.~Catani, G.~Ferrera and M.~Grazzini,
  \emph{$W$ boson production at hadron colliders: the lepton charge asymmetry in NNLO QCD,}
  {\it JHEP} {\bf 1005} (2010) 006
  [{\tt arXiv:1002.3115}].


\end{thebibliography}
\end{document}